\documentclass[a4paper,10pt]{article}
\usepackage[top=3.5cm,right=2cm,left=2cm]{geometry}
\usepackage[T1]{fontenc} % codifica dei font
\usepackage[utf8]{inputenc} % lettere accentate da tastiera
\usepackage[english]{babel} % lingua del documento
\usepackage{url} % per scrivere gli indirizzi Internet
\usepackage{amsmath}
\usepackage{amsfonts}
\usepackage{cancel}
\usepackage{caption}
\usepackage{wasysym}
\usepackage{widetext}
\usepackage{multicol}
\usepackage{amsthm}
\usepackage{float}
\usepackage{pgf,tikz}
\usepackage{amssymb}
\usepackage{xcolor}
\usepackage{lipsum}
\newenvironment{Figure}
{\par\medskip\noindent\minipage{\linewidth}}
{\endminipage\par\medskip}
\usepackage{graphicx}
\usepackage{multicol}
\date{}

\title{\large{\textbf{Polymer representation of the Bianchi IX Cosmology in the Misner variables}}}
\author{\normalsize{Eleonora Giovannetti,$^1$}\quad\normalsize{Giovanni Montani$^{1,2}$} \\  
	\small{$^1$\emph{Physics Department, “Sapienza” University of Rome, P.le A. Moro 5 (00185) Roma, Italy}} \\
	\small{$^2$\emph{ENEA, Fusion and Nuclear Safety Department, C.R. Frascati, via E. Fermi 45, 00044 Frascati (Roma), Italy}}}
\begin{document}
	
\maketitle 

\begin{abstract}
We analyze the Bianchi IX Universe in the Polymer Quantum Mechanics framework by facing both semiclassical and purely quantum effects near the cosmological singularity. We adopt Misner variables to describe the model dynamics, applying the polymer paradigm simultaneously to the isotropic one (linked to the Universe volume) and to the two anisotropy ones (characterizing the physical gravitational degrees of freedom). Setting two different cut-off scales for the two different variable sets, i.e. the geometrical volume and the gravity tensor modes, we demonstrate how the semiclassical properties of the Bianchi IX dynamics are sensitive to the ratio of the cut-off parameters. In particular, the semiclassical evolution turns  out to be chaotic only if the parameter associated to the volume discretization is greater or equal to that one of the anisotropies. Concerning the chaotic case we perform a purely quantum polymer analysis, demonstrating that the original Misner result about the existence of quasi-classical states near the singularity (in the sense of high occupation numbers) is still valid in the revised approach and able to account for cut-off physics effects. The possibility for a comparison with the original study by Misner is possible because the singularity is still present in the semiclassical 
evolution of the cosmological model for all the parameter space. 
We interpret this surprising feature as the consequence of a geometrical volume discretization which does not prevent the volume from vanishing, i.e. restoring in the Minisuperspace analysis its zero value.% insert abstract here
\end{abstract}

\begin{multicols}{2}
	
\section{Introduction}

One of the most puzzling questions concerning the viability of General Relativity as a physical theory consists in its feature of predicting singularities of the space-time, where the radial force and the matter-energy density diverge, especially in the cosmological implementations \cite{PH1,PH2,PC}. The idea that the classical theory must be abandoned, in the limit of extreme curvature values, in view of the inclusion of quantum effects has been not satisfactory addressed by the canonical approach based on the 
Wheeler-DeWitt (WDW) equation \cite{DW,I,PC,CQG}. 
In this scenario, the volume of the Universe (more in general, the metric determinant) seems to have the properties of an internal time-like variable for the theory and therefore the emergence of a Big Bounce
is conceptually forbidden by the natural flows 
of a clock \cite{QC,CQG}. 

Loop Quantum Gravity theory \cite{R1,R2,T,CQG} had more success. Indeed, it was very successful in determining a Bouncing Cosmology thanks to the discrete nature of the volume and the graph structure of the theory \cite{A1,A2,A3}, altough its cosmological application suffers some limitations with respect to the survival of the $SU(2)$ symmetry in the Minisuperspace formulation \cite{Cianfrani1,Cianfrani2}.

A more phenomenological way to investigate the effects of including cut-off physics in the cosmological dynamics has been identified \cite{Corichi} in the 
implementation of the Polymer Quantum Mechanics to the cosmological setting, both on a semiclassical and on a purely quantum dynamics \cite{S,O,M,Chiara,Moriconi,Battisti}. Such a reformulation of Canonical Quantum Cosmology is able to bring some physical features of the discretized geometry into the dynamical problem, avoiding some technical difficulties that Loop Quantum Cosmology finds in treating general enough cosmological models, like the Bianchi IX Universe (the most general allowed by the homogeneity constraint) \cite{QC,MU,G,PC}. 
In \cite{O} it has been studied the implementation of the Polymer Quantum Mechanics to the anisotropy Misner variables of a Bianchi IX Universe, demonstrating how the semiclassical evolution can no longer be chaotic. In \cite{Chiara} the same study has been 
performed by polymer quantizing the Universe volume	alone, i.e. the standard isotropic Misner variable. This study is still associated to the presence of semiclassical chaos and the quantum features derived by Misner about quasi-classical states surviving near the singularity are similarly reproduced. The singularity is still present in both these analyses, in contrast with the general result derived in \cite{S}, where the polymer paradigm is implemented to the cubed scale factor variable instead the usual Misner one, and the non-singular Cosmology predicted in the Loop Quantum Cosmology framework (see \cite{WE1,WE2,WE3}). In particular, in \cite{WE2} it is shown that the cosmological singularity of the Bianchi IX space-time is replaced by a bounce that can be approximated as an instantaneous 
transition between two classical Bianchi I solutions, with simple transition rules if the relation between the solutions 
before and after the bounce is expressed in terms of the Misner variables. The reconciling point of view of these discrepancies is the observation that the zero eigenvalue of the volume is naturally suppressed in the model quantum dynamics when the cubed scale factor is polymer discretized. Conversely, when the standard Misner variable is considered, i.e. the logarithm of the natural scale factor, its discretization does not prevent the semiclassical emergence of an asymptotic zero value for the Universe volume. 

The present study develops such a point of view to its maximal extent by requiring that the Polymer Quantum Mechanics is applied to all the Minisuperspace variables, disregarding their physical nature. Indeed, there is no well-grounded reason to claim that in a quantum gravity approach the Universe volume and its anisotropies require a separate treatment (like it is done in the so-called multi-time approach \cite{IK1,IK2}), above all in order to do not violate the geometrical nature of the gravitational field. We adopt two different polymer scale parameters for the isotropic and the two anisotropic Misner variables, 
in order to better characterize the relative influence of the two sets of variables and to make easier the comparison with previous approaches. 

We find that the chaotic properties of the semiclassical Bianchi IX dynamics survive only as far as the isotropic variable discretization scale is equal 
or greater than the anisotropic variables one. As soon as the ratio of the two scales is less than one, the features of the original analysis in \cite{O} are clearly recovered and no quasi-classical occupation numbers can be obtained in the full quantum treatment near the initial singularity. On the contrary, when the ratio is greater than (or equal to) one, 
i.e. when the chaos of the model is preserved, the quantum analysis can be developed according to the same scheme proposed by Misner in \cite{QC} and the quasi-classical states still emerge. For completeness, it has been discussed the validity of the quantum results in the absence of the square approximation that Misner used to solve the eigenvalue problem during the quantum treatment of the model and a brief analysis of the exact triangular problem in the polymer formulation has been developed.

Apart from its technical content, the main merit of this investigation is demonstrating how in the polymer representation the chaotic features of the Bianchi IX Cosmology depend on the reciprocal relevance of the cut-off scales concerning the Universe volume and the gravitational degrees of freedom. Furthermore, we demonstrate that a full polymer quantization of all the Minisuperspace variables does not remove the singularity, simply because the discretization of the isotropic Misner variable maintains the possibility of a vanishing value of the space volume.

\section{The polymer representation of quantum mechanics}
The Polymer Quantum Mechanics is an alternative representation of the canonical commutation relations, non-equivalent to the usual Schr\"{o}dinger one. It is based on the assumption that one or more variables of the phase space are discretized. Therefore, it can be used to investigate the consequences of the introduction of a physical cut-off. 

\subsection{Polymer kinematics}

In order to introduce the polymer representation without any reference to the Schr\"{o}dinger one, let us consider the abstract kets $|\mu\rangle$ labelled by the real parameter $\mu\in\mathbb{R}$ and taken from the Hilbert space $\mathcal{H}_{poly}$. 

A generic state can be defined through a finite linear combination of them:

\begin{equation}
|\psi\rangle=\sum_{i=1}^Na_i|\mu_i\rangle\,,
\end{equation}

where $\mu_i\in\mathbb{R},\;i=1,\dots,N\in\mathbb{N}$.

We choose the inner product so that the fundamental kets are orthonormal:

\begin{equation}
\langle\mu|\nu\rangle=\delta_{\mu\nu}\,.
\end{equation}

It can be demonstrated that $\mathcal{H}_{poly}$ is non-separable. 

There are two fundamental operators that can be defined on this Hilbert space: the symmetric \emph{label operator} $\hat{\epsilon}$ and the unitary \emph{shift operator} $\hat{s}(\lambda)$ with $\lambda\in\mathbb{R}$. They act on the kets $|\mu\rangle$ as follows: 

\begin{equation}
\hat{\epsilon}|\mu\rangle:=\mu|\mu\rangle\,;\quad\hat{s}(\lambda)|\mu\rangle:=|\mu+\lambda\rangle\,.
\end{equation}

Since the kets $|\mu\rangle$ and $|\mu+\lambda\rangle$ are orthogonal $\forall\lambda$, the \emph{shift operator} $\hat{s}(\lambda)$ is discontinuous in $\lambda$ and there is no Hermitian operator that can generate it by exponentiation. 

Now we consider a one-dimensional system $(q,p)$ and we denote the wave functions in the $p$-polarization as $\psi(p):=\langle p|\psi\rangle$, where $\psi_{\mu}(p):=\langle p|\mu\rangle = e^{i\mu p}$. 

It is easy to see that the \emph{shift operator} acts in this polarization as

\begin{equation}
\hat{s}(\lambda)\cdot\psi_{\mu}(p)= e^{i\lambda p}e^{i\mu p}=e^{i(\mu+\lambda) p}=\psi_{\mu+\lambda}(p)\,,
\end{equation}

so $\hat{s}(\lambda)$ can be identified with the operator $e^{i\lambda\hat{p}}$ but $\hat{p}$ cannot be defined rigorously. 

On the other hand, the operator $\hat{q}$ acts as a differential operator in this polarization and therefore it corresponds to the \emph{label operator} $\hat{\epsilon}$:

\begin{equation}
\hat{q}\cdot\psi_{\mu}(p)= -i\frac{\partial}{\partial p} \psi_{\mu}(p)= \mu \psi_{\mu}(p)\,.
\end{equation}

It can be noticed that the eigenvalues of $\hat{q}$ can be considered as a discrete set because they label kets that are always orthonormal.  

\subsection{Polymer dynamics}{\label{Polymer}}

After introducing the polymer kinematical Hilbert space, we need to solve the problem related to the definition of the physical operators. 

Let us consider a one-dimensional system described by the Hamiltonian

\begin{equation}
H=\frac{p^2}{2m}+\mathcal{V}(q)
\end{equation}

in the $p$-polarization. As seen in the previous section, if we assume that $\hat{q}$ is a discrete operator, we have to find an approximate form for $\hat{p}$. The standard procedure (widely described in \cite{Corichi}) is to introduce a regular lattice with spacing $\mu_0$:

\begin{equation}
\gamma_{\mu_0}=\{q \in \mathbb{R} : q=n\mu_0, \; \forall\;  n \in \mathbb{Z}\}\,.
\end{equation}

In order to remain in the lattice, the only states permitted are $|\psi\rangle=\sum_{n}b_n|\mu_n\rangle\in\mathcal{H}_{\gamma_{\mu_0}}$ where $\mu_n=n\mu_0$. Also, it is possible to use the action of the operator $e^{i\lambda\hat{p}}$, after it has been restricted to the lattice, to define an approximate version of $\hat{p}$:

\begin{eqnarray}
\nonumber\hat{p}_{\mu_0}|\mu_n\rangle&&=\frac{1}{2i\mu_0}[e^{i\mu_0\hat{p}}-e^{-i\mu_0\hat{p}}]|\mu_n\rangle=\\&&=\frac{1}{2i\mu_0}(|\mu_{n+1}\rangle-|\mu_{n-1}\rangle)\,.
\end{eqnarray}

It derives from the fact that for $p\ll\frac{1}{\mu_0}$ one gets $p\backsimeq\frac{1}{\mu_0}\sin(\mu_0p)=\frac{1}{2i\mu_0}(e^{i\mu_0p}-e^{-i\mu_0p})$.

Now it is possible to introduce an approximate version of $\hat{p}^2$:

\begin{eqnarray}
\nonumber\hat{p}^2_{\mu_0}|\mu_n\rangle&&\equiv\hat{p}_{\mu_0}\cdot\hat{p}_{\mu_0}|\mu_n\rangle=\\\nonumber&&=\frac{1}{4\mu_0^2}[-|\mu_{n-2}\rangle+2|\mu_{n}\rangle-|\mu_{n+2}\rangle]=\\&&=\frac{1}{\mu_0^2}\sin^2(\mu_0p)|\mu_n\rangle\,.
\end{eqnarray}

We remind that $\hat{q}$ is a well-defined operator, so the regularized version of the Hamiltonian writes as

\begin{equation}
\hat{H}_{\mu_0}:=\frac{1}{2m}\hat{p}_{\mu_0}^2+\hat{V}(q)
\end{equation}

and represents a symmetric and well-defined operator on $\mathcal{H}_{\gamma_{\mu_0}}$.

\section{The Mixmaster model in the Misner variables}

In this section we introduce the most important results about the semiclassical and quantum features of the Bianchi IX dynamics towards the cosmological singularity. This model was named Mixmaster by Misner and it is the most general homogeneous space which admits an isotropic limit.

The line element in the Misner variables \cite{MU} for this model is

\begin{equation}
ds^2=-N(t)^2dt^2+\frac{1}{4}e^{2\alpha}(e^{2\beta})_{ij}\sigma_i\sigma_j\,,
\end{equation}

where $N(t)$ is the \emph{lapse function}, $\sigma_i$ are three differential forms depending on the Euler angles of the SO(3) group of symmetry and $\beta$ is a diagonal, traceless matrix. Therefore, it can be parametrized in terms of the two independent variables $(\beta_+,\beta_-)$ as $\beta_{ij}=diag(\beta_++\sqrt{3}\beta_-,\beta_+-\sqrt{3}\beta_-,-2\beta_+)$. The Misner variables $(\alpha,\beta_\pm)$ represent a very convenient choice to parametrize the line element. In particular, the variable $\alpha$ is that one related to the volume and the expansion of the Universe ($\mathcal{V}\sim e^{3\alpha}$, $\mathcal{V}$ denoting a fiducial volume), while $(\beta_+,\beta_-)$ contain all the information about the anisotropies and the shape deformations. Moreover, the Hamiltonian of the model results to be very simple when it is written in the Misner variables because its kinetical term is diagonal. Then, we can write the super-Hamiltonian as follows:

\begin{equation}
\label{H}
\mathcal{H}=\frac{N\kappa}{3(8\pi)^2}e^{-3\alpha}\bigg(-p_{\alpha}^2+p_+^2+p_-^2+\frac{3(4\pi)^4}{\kappa^2}e^{4\alpha}V(\beta_\pm)\bigg)=0\,,
\end{equation}

where $(p_\alpha,p_\pm)$ are the conjugate momenta to $(\alpha,\beta_\pm)$ respectively and $\kappa=8\pi G$ (we are in $c=1$ units). Due to the spatial curvature, $V(\beta_\pm)$ is a potential term depending only on the anisotropy variables, whose explicit form is the following:

\begin{multline}
\label{V}
V(\beta_\pm)=2e^{4\beta_+}\big(\cosh(4\sqrt{3}\beta_-)-1	\big)+\\-4e^{2\beta_+}\cosh(2\sqrt{3}\beta_-)+e^{-8\beta_+}\,.
\end{multline}

The potential walls are steeply exponential and define a closed domain with the symmetry of an equilateral triangle, as we can see in Fig. \ref{fig:3}. These walls move outwards while approaching the cosmological singularity ($\alpha\rightarrow-\infty$) due to the term $e^{4\alpha}$ in \eqref{H} (which increases towards the singularity, i.e. $\alpha\rightarrow-\infty$).

\begin{Figure}[h!]
	\centering
	\includegraphics[width=0.8\linewidth]{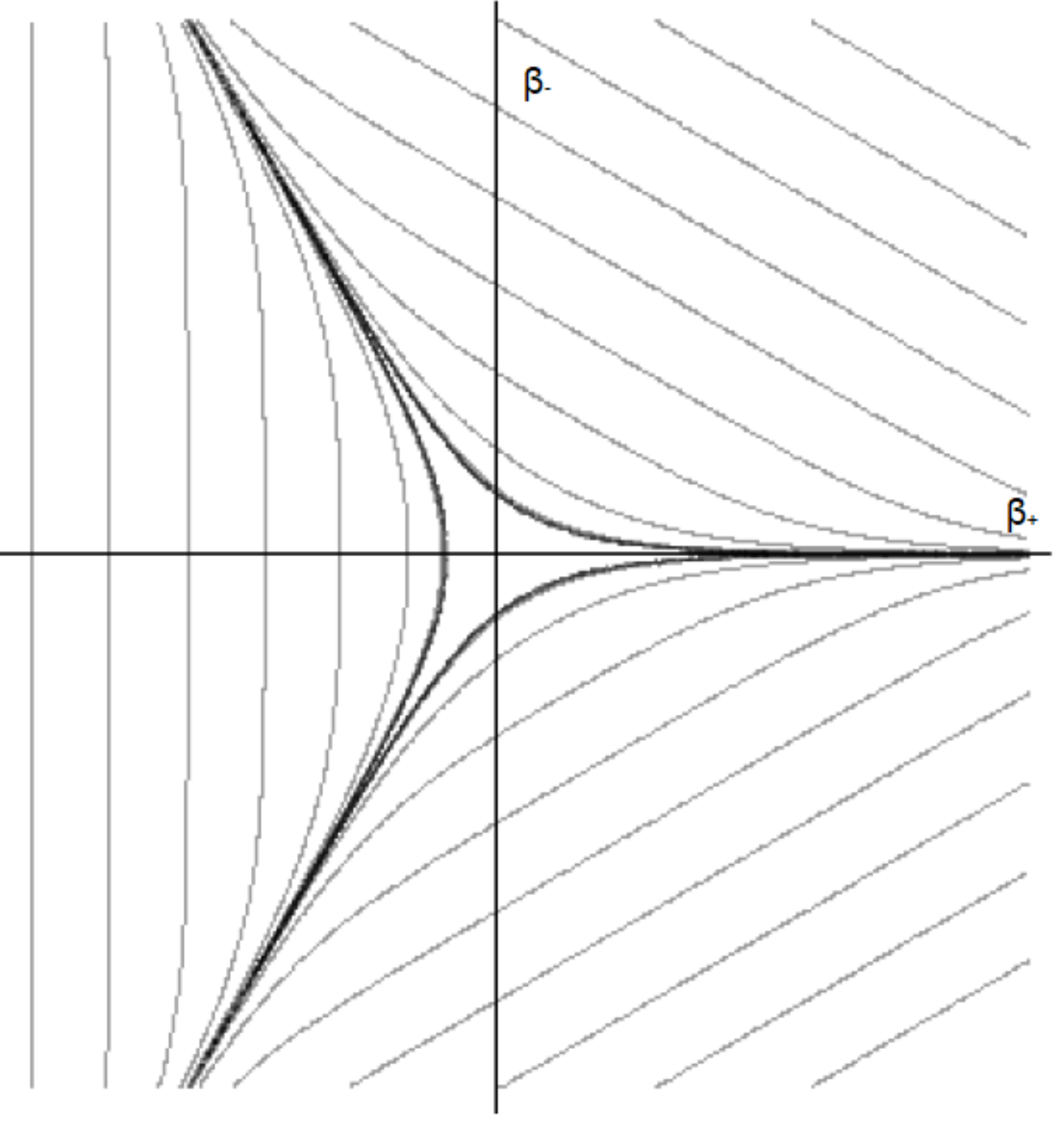}
	\captionof{figure}{\label{fig:3}Equipotentials of the function $V(\beta_\pm)$ in the $(\beta_+,\beta_-)$-plane.}
\end{Figure}

Now it is possible to apply the Arnowitt-Deser-Misner (ADM) reduction \cite{ADM} 
to connect the study of the Mixmaster dynamics with that one of a pinpoint particle (called point-Universe) that moves in a potential well. This procedure consists in solving the super-Hamiltonian constraint with respect to the momentum conjugated to the new time coordinate of the phase space, which here is identified with $\alpha$. Then, one obtains the so-called \emph{reduced ADM-Hamiltonian}

\begin{equation}
\label{HADM}
H_{ADM}:=-p_{\alpha}=\sqrt{p_+^2+p_-^2+e^{4\alpha}\frac{3(4\pi)^4}{\kappa^2}V(\beta_\pm)}\,.
\end{equation}

The classical dynamics of the model can be analyzed by studying the corresponding Hamilton's equations that describe the evolution (in function of $\alpha$) of the point-Universe, identified by the  coordinates ($\beta_+,\beta_-$), moving in the potential well. Because of the steepness of the walls, we can consider the point-Universe as a free particle for most of its motion, except when it occurs a bounce against one of the three walls. Using the free-particle approximation ($V(\beta_\pm)\sim 0$), it is possible to derive the velocity of the point-Universe from the Hamilton's equations, that is called \emph{anisotropy velocity}:

\begin{equation}
\label{beta}
\beta'\equiv\sqrt{{\beta'_+}^2+{\beta'_-}^2}=1\,.
\end{equation}

On the other hand, the study of the region in which the potential term becomes relevant gives information about the position of the potential wall and shows that it moves outwards with velocity $|\beta'_{wall}|=\frac{1}{2}$, so a bounce is always possible. In particular, every bounce occurs according to the following reflection law:

\begin{equation}
\label{legge}
\frac{1}{2}\sin(\theta_i+\theta_f)=\sin\theta_i-\sin\theta_f\,,
\end{equation}

where $\theta_i$ and $\theta_f$ are the
incidence angle and the reflection one taken with respect to the normal of the potential wall. Moreover, the condition for a bounce to occur is represented by the existence of a maximum incidence angle

\begin{equation}
\theta_{max}\equiv\arccos\bigg(\frac{1}{2}\bigg)=\frac{\pi}{3}\,.
\end{equation}

This result, together with the triangular symmetry of the system, shows that the point-Universe always has a bounce against one of the three potential walls. 

In conclusion, the ADM reduction procedure in the Misner parametrization maps the dynamics of the Mixmaster Universe into the motion of a pinpoint particle inside a closed two-dimensional domain.  The particle undergoes an infinite series of bounces against the potential walls while approaching the singularity and the motion between two subsequent bounces is a uniform rectilinear one. The only effect of a bounce is to change the direction of the particle in a way that, as it goes towards the singularity, it will assume all the possible directions, regardless the initial condition, and so it will experience a chaotic dynamics. 

Another important benefit of using the Hamiltonian formalism is that it allows to quantize the system in the canonical way. As we are interested in the features of the model near the cosmological singularity, we have to investigate the quantum effects that can modify the classical dynamics. Following the canonical formalism, it is possible to solve the quantum problem that corresponds to find the eigenvectors and the eigenvalues of a pinpoint particle in a two-dimensional potential well. 

Taking advantage of the geometric properties of the system, Misner obtained the following \emph{adiabatic invariant} which represents the most important result of his quantum analysis:

\begin{equation}
\label{const}
\big<H_{ADM}\alpha\big>=const\,,
\end{equation}

where the symbol $\big<\dots\big>$ denotes the average value over a large number of bounces. It is possible to use this sort of conservation law to study the behaviour of the quasi-classical states (i.e. quantum states with very high occupation numbers) towards the singularity. In particular, as shown in \cite{QC}, using the quasi-classical approximation and taking the limit $\alpha\rightarrow-\infty$, the expression \eqref{const} becomes

\begin{equation}
\big<m^2+n^2\big>=const\,,
\end{equation}

where $m,n$ are the positive and integer quantum numbers related to the anisotropies $(\beta_+,\beta_-)$.

Therefore, we can conclude that if the present Universe is in a quasi-classical state of anisotropy ($n^2+m^2\gg1$), then it must have been in the same quasi-classical state also near the cosmological singularity. 

\section{Semiclassical analysis of the Mixmaster model in the polymer representation}
\label{Semiclassical}

The aim of this section is to discuss the main features of the Mixmaster semiclassical dynamics in the polymer representation. The word \textquotedblleft semiclassical" means that the polymer-modified super-Hamiltonian constraint \eqref{super} is obtained as the lowest order term of a WKB expansion for $\hbar\rightarrow0$. 

We choose to define the Misner variables $(\alpha,\beta_\pm)$ as discrete ones, so we have to find an approximated form for the operators $(p_\alpha^2,p_\pm^2)$, as seen in Sec. \ref{Polymer}. This procedure consists in the following formal substitutions:
\begin{align}
\label{pol1}
p_\pm^2&\rightarrow\frac{1}{\mu^2}\sin^2(\mu p_\pm)\,, \\
\label{pol2}
p_\alpha^2&\rightarrow\frac{1}{\mu_\alpha^2}\sin^2(\mu_\alpha p_\alpha)\,,
\end{align}

where $\mu$ is the polymer parameter for the anisotropies, while $\mu_\alpha$ is that one related to the isotropic variable $\alpha$.

The super-Hamiltonian constraint \eqref{H} becomes

\begin{eqnarray}
\label{super}
\mathcal{H}^{poly}&&=\frac{N\kappa}{3(8\pi)^2}e^{-3\alpha}\bigg[-\frac{1}{\mu_\alpha^2}\sin^2(\mu_\alpha p_\alpha)+\\\nonumber&&+\frac{1}{\mu^2}\sin^2(\mu p_+)+\frac{1}{\mu^2}\sin^2(\mu p_-)+\frac{\mathcal{V(\beta_\pm)}}{\mu^2_\alpha}\bigg]=0\,,
\end{eqnarray}

while the reduced Hamiltonian has the following expression:

\begin{equation}
\label{Ham}
H_{\alpha}=\frac{1}{\mu_\alpha}\arcsin\sqrt{\frac{\mu_\alpha^2}{\mu^2}[\sin^2(\mu p_+)+\sin^2(\mu p_-)]+\mathcal{V(\beta_\pm)}}\,,
\end{equation}

with the condition $0\leq \frac{\mu_\alpha^2}{\mu^2}[\sin^2(\mu p_+)+\sin^2(\mu p_-)]+\mathcal{V(\beta_\pm)}\leq1$ due to the presence of the arcsin function. 

In both \eqref{super} and \eqref{Ham} we have done the substitution $\mathcal{V}(\beta_\pm)=\mu_\alpha^2\frac{3(4\pi)^4e^{4\alpha}}{\kappa^2}V(\beta_\pm)$.

The dynamics of the model is described by the Hamilton's equations, obtained from the reduced Hamiltonian shown in \eqref{Ham}:

\begin{subequations}
	\label{EqHam}
	\begin{align}
	\label{b'}
	\beta_\pm'&=\frac{\partial H_{\alpha}}{\partial p_\pm}=\frac{\mu_\alpha}{\mu}\frac{\sin(2\mu p_\pm)}{\sin(2\mu_\alpha H_{\alpha})}
	\\
	\label{p+}
	p_\pm'&=-\frac{\partial H_{\alpha}}{\partial \beta_\pm}=-\frac{3\mu_\alpha(4\pi)^4e^{4\alpha}}{\kappa^2\sin(2\mu_\alpha H_{\alpha})}\frac{\partial V(\beta_\pm)}{\partial\beta_\pm}
	\\
	\label{Halpha}
	H_\alpha'&=4e^{4\alpha-8\beta_+}\frac{3\mu_\alpha(4\pi)^4}{\kappa^2\sin(2\mu_\alpha H_\alpha)}
	\end{align}
\end{subequations}

where we have used the following compact notation

\begin{multline}
\sin(\mu_\alpha H_\alpha)=\\=\sqrt{\frac{\mu_\alpha^2}{\mu^2}[\sin^2(\mu p_+)+\sin^2(\mu p_-)]+\mathcal{V(\beta_\pm)}}
\end{multline}

and the symbol $'$ to denote the derivative with respect to $\alpha$.

According to the discussion of the previous section, we start our analysis studying the system in the case $V(\beta_\pm)=0$. Now we can easily show that the cosmological singularity is still present. In fact, under this condition it can be demonstrated that there is a logarithmic relation between the time variable $t$ and the isotropic one $\alpha$, so we have that $\alpha\sim\ln(t)\rightarrow-\infty$ for $t\rightarrow 0$ even if $\alpha$ is described in the polymer formulation. This result points out that the discrete nature of the isotropic variable $\alpha$ does not prevent the Universe volume from vanishing.

As we consider the point-Universe as a free particle, we have that $(p_+,p_-)$ (and consequently $H_\alpha$) are constants of motion and the modified anisotropy velocity has the following expression:

\begin{eqnarray}
\label{velocity}
\beta'&&\equiv\sqrt{\beta_+'^2+\beta_-'^2}=\\\nonumber&&=\sqrt{\frac{\sin^2(\mu p_+)\cos^2(\mu p_+)+\sin^2(\mu p_-)\cos^2(\mu p_-)}{\Delta^2(\mu p_+,\mu p_-)\big[1-\frac{\mu_\alpha^2}{\mu^2}\Delta^2(\mu p_+,\mu p_-)\big]}}\,,
\end{eqnarray}

where $\Delta(x,y)=\sqrt{\sin^2(x)+\sin^2(y)}$.  
It is easy to realize that, taking $\mu$ and $\mu_\alpha$ vanishing, the relation \eqref{beta} is recovered.

To verify if the dynamics has the same chaotic features of the standard case, it is necessary to study the relative motion between the particle and the potential walls, whose velocity is still $|\beta'_{wall}|=\frac{1}{2}$ because the polymer representation leaves the potential unchanged. On the other hand, if we study the anisotropy velocity of the particle while varying the values of the polymer parameters, we can see that

\begin{equation}
\label{caos}
\beta'\equiv r(\mu_\alpha, \mu, p_\pm)\geq1\;\forall\;p_\pm\in\mathbb{R}\Leftrightarrow\frac{\mu_\alpha}{\mu}\geq1\,.
\end{equation}

Therefore, we can conclude that the dynamics of the Mixmaster model is expected to be still chaotic under this condition, because of the existence of the singularity and the presence of a never-ending series of bounces against the potential walls. Instead, if we choose the polymer parameters such that $\mu_\alpha/\mu<1$ (Fig. \ref{fig:5}), the series of bounces occurs until the particle velocity becomes smaller than the velocity of the potential walls and then the point-Universe reaches the singularity with no other bounces. The figures represented in Fig. \ref{fig:4} show that, by choosing the ratio $\mu_\alpha/\mu$  greater than or equal to one, the anisotropy velocity is always above the plane $z=1$ (here $z$ is the vertical coordinate).

In order to derive the modified reflection law in the polymer representation, we introduce the following parametrization for the anisotropy velocity:

\begin{eqnarray}
\label{par}
\nonumber&&(\beta'_-)_i=r_i\sin\theta_i\quad\;\;\,(\beta'_-)_f=r_f\sin\theta_f
\\&&
(\beta'_+)_i=-r_i\cos\theta_i\;\;\;(\beta'_+)_f=r_f\cos\theta_f
\end{eqnarray}

\begin{Figure}[h!]
	\centering
	\includegraphics[width=1\linewidth]{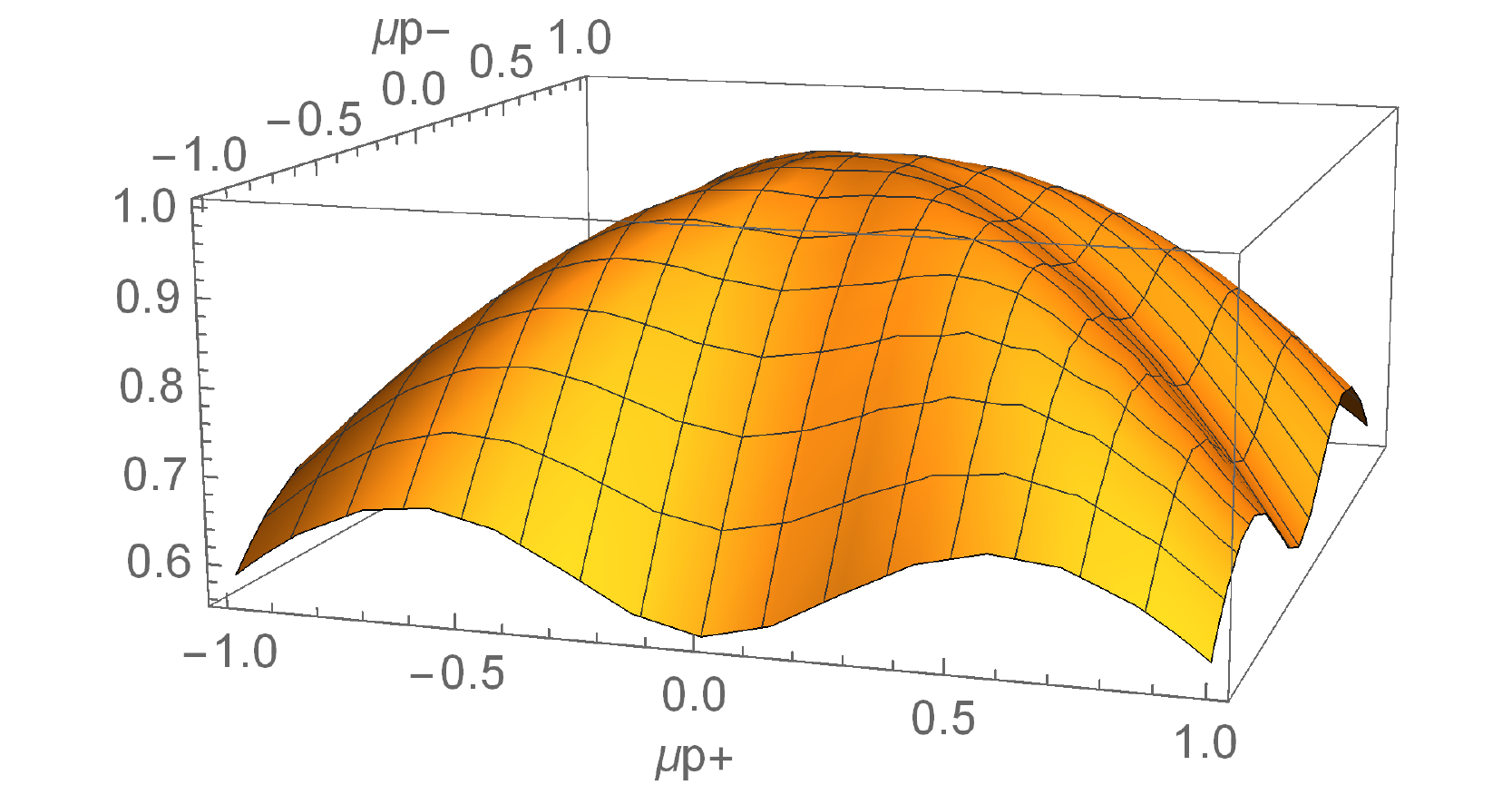}
	\captionof{figure}{\label{fig:5} 3D-profile of the anisotropy velocity \eqref{velocity} with $\mu_\alpha/\mu=0.1$.}
\end{Figure} 

\begin{Figure}[h!]
	\centering
	\includegraphics[width=0.9\linewidth]{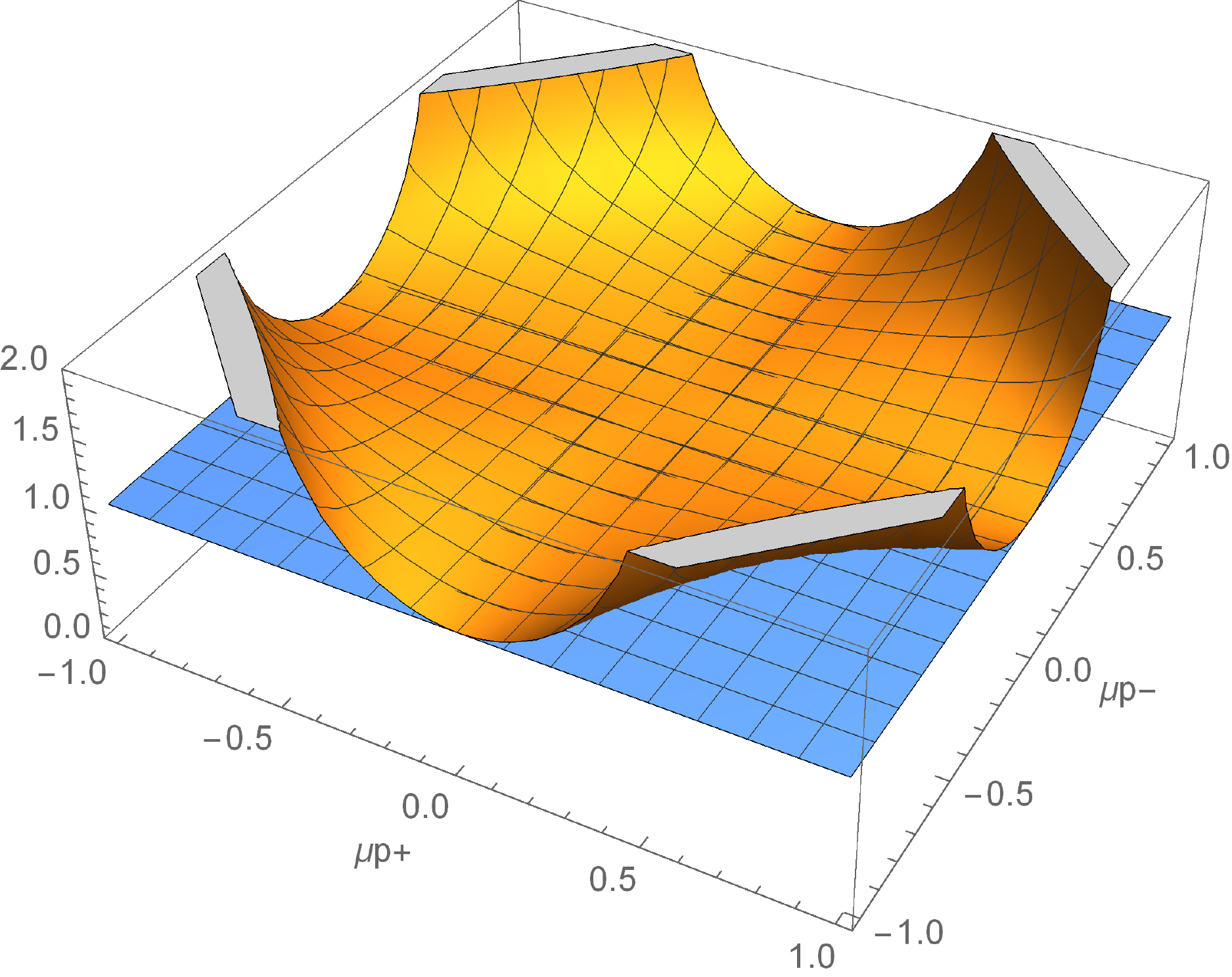}\quad\includegraphics[width=0.9\linewidth]{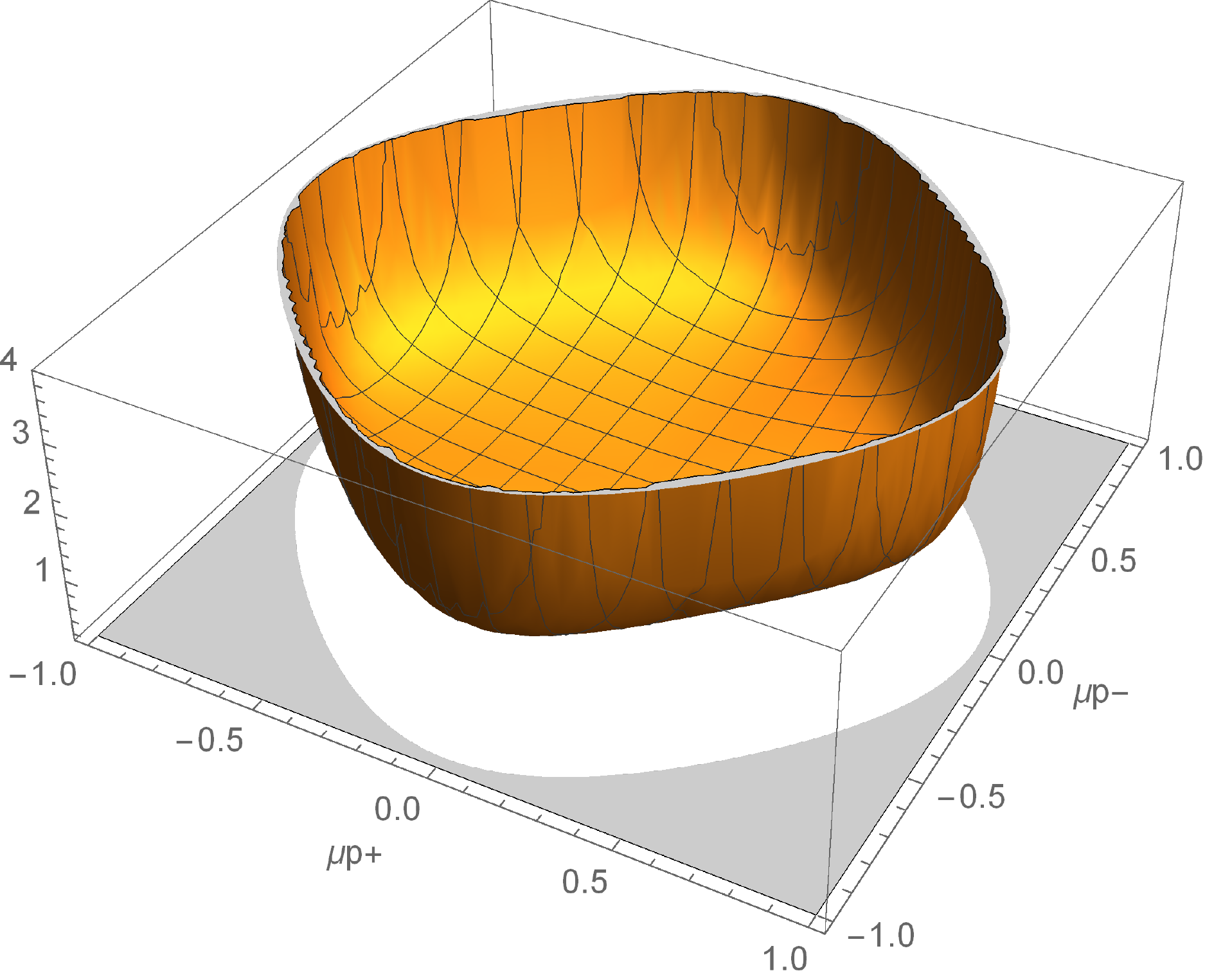}
	\captionof{figure}{\label{fig:4} 3D-profiles of the anisotropy velocity \eqref{velocity} with $\mu_\alpha/\mu=1$ in the first figure and $\mu_\alpha/\mu=1.5$ in the second one.}
\end{Figure}

where $\theta_i$ and $\theta_f$ are respectively the incident and the reflection angles and $r_i$ and $r_f$ are the particle velocities before and after the bounce\footnote{Once the particle has bounced against one of the walls, the values assumed by the constants of the free-particle motion change and so the value of the anisotropy velocity changes \textbf{too}.}.

The condition for a bounce to occur is $(\beta'_+)_i>\beta_{wall}$ and this led to the existence of a maximum incidence angle

\begin{equation}
\theta_i<\theta^{poly}_{max}=\arccos\bigg(\frac{1}{2r_i}\bigg)\,.
\end{equation}

Thanks to this result, we can confirm our analysis about the chaotic features of the dynamics. Indeed, if $\mu_\alpha/\mu<1$ ($\Rightarrow r_i<1$) we find $\theta^{poly}_{max}<\frac{\pi}{3}$ and this imply the absence of bounces even if the particle moves towards a specific wall. On the other hand, if $\mu_\alpha/\mu\geq1$ ($\Rightarrow r_i\geq1$) we have $\frac{\pi}{3}\leq\theta^{poly}_{max}<\frac{\pi}{2}$ and this means that a bounce is always possible, given the triangular symmetry of the system. 

Now, in order to find the relation between the particle directions before and after a bounce, we have to find the constants of motion when the point-Universe is nearby the potential walls. In particular, thanks to the symmetry of the system, we can consider only one of the three walls and use the potential term $V(\beta_\pm)=e^{-8\beta_+}$. In these conditions, the constants of motion are the following:

\begin{align}
K_1&=p_-\,,\\
K_2&=H_\alpha-p_+/2\,.
\end{align}

The first one comes from the fact that the Hamiltonian $H_\alpha$ depends only from $\beta_+$, while the second one derives from the Hamilton's equations \eqref{p+} and \eqref{Halpha}. 
We use the constants of motion to write the two following conservation laws:

\begin{equation}
\label{p-}
r_i\sin\theta_i\sin(2\mu_\alpha H_{\alpha_i})=r_f\sin\theta_f\sin(2\mu_\alpha H_{\alpha_f})
\end{equation}
\begin{multline}
\label{Hcost}
H_{\alpha_i}-\frac{1}{4\mu}\arcsin\bigg[-\frac{\mu}{\mu_\alpha}r_i\cos\theta_i\sin(2\mu_\alpha H_{\alpha_i})\bigg]=\\=H_{\alpha_f}-\frac{1}{4\mu}\arcsin\bigg[\frac{\mu}{\mu_\alpha}r_f\cos\theta_f\sin(2\mu_\alpha H_{\alpha_f})\bigg]
\end{multline}

obtained by using the parametrization \eqref{par} and an explicit expression for $(p_+,p_-)$ derived by the Hamilton's equations \eqref{p+}. 

After the substitution of \eqref{p-} in \eqref{Hcost} and the use of the explicit expressions\footnote{In using the expression \eqref{Ham} we have neglected the potential term $\mathcal{V(\beta_\pm)}$ because we are sufficiently far from the region where the potential becomes relevant.} \eqref{velocity} for $r_{i,f}$ and \eqref{Ham} for $H_\alpha$ respectively, we are able to derive the following reflection law:

\begin{eqnarray}
\nonumber\frac{1}{4\mu}\bigg[\arcsin\bigg(\cos\theta_f\frac{\sin\theta_i}{\sin\theta_f}\Delta(2\mu p_+^i,2\mu p_-^i)\bigg)&&+\\\nonumber+\arcsin\bigg(\cos\theta_i\Delta(2\mu p_+^i,2\mu p_-^i)\bigg)\bigg]&&=\\\nonumber=\frac{1}{\mu_\alpha}\bigg[\arcsin\bigg(\frac{\mu_\alpha}{\mu}\Delta(\mu p_+^f,\mu p_-^f)\bigg)&&+\\-\arcsin\bigg(\frac{\mu_\alpha}{\mu}\Delta(\mu p_+^i,\mu p_-^i)\bigg)&&\bigg]\,,
\end{eqnarray}

where $\Delta(x,y)=\sqrt{\sin^2(x)+\sin^2(y)}$.

In order to make a comparison with the standard case \eqref{legge}, an expansion up to the second order for $\mu$ and $\mu_\alpha$ is required. The final result is:

\begin{equation}
\label{reflection}
\frac{1}{2}\sin(\theta_i+\theta_f)=\sin\theta_i(1+\Pi_f^2)(1+R_f)-\sin\theta_f(1+\Pi_i^2)(1+R_i)
\end{equation}

where

\begin{equation}
\begin{cases}
\Pi^2=\frac{1}{6}\mu_\alpha^2p^2
\\
R=\frac{2}{3}\mu^2\frac{(p_+)^4+(p_-)^4}{(p_+)^2+(p_-)^2}
\end{cases}
\end{equation}

and it is easy to show that in the limit $\mu,\mu_\alpha\rightarrow0$ we find the standard reflection law \eqref{legge} obtained by Misner \cite{QC}.

Recovering this limit leads us to infer that the map above still admits stochastic properties and therefore, when $\mu_\alpha<\mu$, sooner or later the parameter region where no bounce takes place is reached, like in \cite{O}.

\section{Quantum properties of the polymer Mixmaster model}

In this section we analyze the quantum properties of the Mixmaster model in the polymer representation by solving the WDW equation, which corresponds to the quantization of the modified super-Hamiltonian constraint \eqref{super}.
The WDW for the polymer Mixmaster model reads as follows:

	\begin{eqnarray}
	\label{WDW}
	&&\bigg[-\frac{1}{\mu_\alpha^2}\sin^2(\mu_\alpha p_\alpha)+\frac{1}{\mu^2}\sin^2(\mu p_+)+\frac{1}{\mu^2}\sin^2(\mu p_-)+\nonumber\\&&+\frac{3(4\pi)^4}{\kappa^2}e^{4\alpha}\hat{V}(\beta_\pm)\bigg]\Psi(p_\alpha,p_\pm)=0
	\end{eqnarray}

where all the canonical variables have been promoted to operators, using the approximate versions for $p_\alpha$ and $p_\pm$. As we are interested in the behaviour of the model near the singularity, we consider the potential walls as perfectly vertical and so we outline the particle motion as that one of a free particle with appropriate boundary conditions (particle in a box). 

We start with imposing $\hat{V}(\beta_\pm)=0$ in \eqref{WDW} and solving the free-particle problem by searching for solutions of the form

\begin{equation}
\Psi(p_\alpha,p_\pm)=\chi(p_\alpha)\phi(p_\pm)\,.
\end{equation} 

The eigenvalue problem for the anisotropy component $\phi(p_\pm)$ of the wave function is the following:

\begin{multline}
\bigg[\frac{1}{\mu^2}\sin^2(\mu p_+)+\frac{1}{\mu^2}\sin^2(\mu p_-)\bigg]\phi(p_\pm)=k^2\phi(p_\pm)
\end{multline}

and can be solved by further separating the variables, i.e.

\begin{equation}
\begin{cases}
\big[\frac{1}{\mu^2}\sin^2(\mu p_+)-k_+^2\big]\phi_+(p_+)=0
\\
\big[\frac{1}{\mu^2}\sin^2(\mu p_-)-k_-^2\big]\phi_-(p_-)=0
\end{cases}
\end{equation}

where $\phi(p_\pm)=\phi_+(p_+)\phi_-(p_-)$ and $k^2=k_+^2+k_-^2$.

It can be easily shown that the eigenfunctions have the following expression:

\begin{equation}
\label{phi}
\begin{cases}
\phi_+(p_+)=A\delta(p_+-p_+^\mu)+B\delta(p_++p_+^\mu)
\\
\phi_-(p_-)=C\delta(p_--p_-^\mu)+D\delta(p_-+p_-^\mu)
\end{cases}
\end{equation}

where $A,B,C,D$ are integration constants and

\begin{equation}
\begin{cases}
p_+^\mu=\frac{1}{\mu}\arcsin(\mu k_+)
\\
p_-^\mu=\frac{1}{\mu}\arcsin(\mu k_-)
\end{cases}
\end{equation}

Now, after the substitution of this result in the WDW-equation for the free particle (\eqref{WDW} with $\hat{V}(\beta_\pm)=0$), we can also solve the problem for the isotropic component of the wave function

\begin{equation}
\bigg(-\frac{1}{\mu_\alpha^2}\sin^2(\mu_\alpha p_\alpha)+k^2\bigg)\chi(p_\alpha)=0
\end{equation}

whose solution is

\begin{equation}
\begin{cases}
\chi(p_\alpha)=\delta(p_\alpha-\bar{p}_\alpha)
\\
\bar{p}_\alpha=\frac{1}{\mu_\alpha}\arcsin(\mu_\alpha k)
\end{cases}
\end{equation}

Does, in order to introduce the boundary conditions for the free-particle motion, we approximate the triangular well with a square one having the same area and vertical walls, in a way that the potential term has the following expression:

\begin{equation}
V(\alpha,\beta_\pm)=\begin{cases} 0 & -\frac{L(\alpha)}{2}\leq\beta_\pm\leq\frac{L(\alpha)}{2} \\ \infty & elsewhere
\end{cases}
\end{equation}

The boundary conditions are imposed on the eigenfunctions \eqref{phi}, expressed in coordinate representation after a Fourier transformation, as follows:

\begin{equation}
\label{phi2}
\begin{cases}
\phi_+\big(\frac{L(\alpha)}{2}\big)=\phi_+\big(-\frac{L(\alpha)}{2}\big)=0\\
\phi_-\big(\frac{L(\alpha)}{2}\big)=\phi_-\big(-\frac{L(\alpha)}{2}\big)=0
\end{cases}
\end{equation}

After solving the system \eqref{phi2}, we are able to write the complete solution for the anisotropy component of the wave function, which has the following expression:

\begin{widetext}
	\begin{equation}
	\label{phi3}
	\phi_{m,n}(\beta_\pm)=\frac{1}{2L(\alpha)}\big(e^{ip_+^\mu\beta_+}-e^{-ip_+^\mu\beta_+}e^{-im\pi}\big)\big(e^{ip_-^\mu\beta_-}-e^{-ip_-^\mu\beta_-}e^{-in\pi}\big)\,,
	\end{equation}
\end{widetext}

where 

\begin{equation}
\label{pmu}
\begin{cases}
p_+^\mu=\frac{m\pi}{L(\alpha)}\\
p_-^\mu=\frac{n\pi}{L(\alpha)}
\end{cases}
\end{equation}

and $m,n\in\mathbb{Z}$ are the quantum numbers associated to the anisotropies $(\beta_+,\beta_-)$. 

At last, the isotropic solution writes as

\begin{equation}
\label{chi}
\begin{cases}
\bar{p}_\alpha=\frac{1}{\mu_\alpha}\arcsin(\mu_\alpha k)
\\
\chi(\alpha)=\int dp_{\alpha} \chi(p_\alpha)e^{ip_\alpha\alpha}=e^{i\bar{p}_\alpha\alpha}=e^{\frac{i}{\mu_\alpha}\int_{0}^{x}dt\arcsin(\mu_\alpha k)}
\end{cases}
\end{equation}

with

\begin{equation}
\label{kappa}
k=\sqrt{\frac{1}{\mu^2}\sin^2\bigg(\mu\frac{m\pi}{L(\alpha)}\bigg)+\frac{1}{\mu^2}\sin^2\bigg(\mu\frac{n\pi}{L(\alpha)}\bigg)}\,.
\end{equation}

It is easy to check that, when $\mu$ and $\mu_\alpha$ are taken vanishing, the Misner solution is recovered and this feature justifies \emph{a posteriori} the implicit assumption to deal with steady potential walls. Indeed, we are recovering the adiabatic approximation which describes the anisotropies quantum dynamics as fast with respect to the potential walls variation.

\subsection{Quasi-classical states near the singularity}

As we stressed in Sec. \ref{Semiclassical}, the Mixmaster semiclassical dynamics preserves its chaotic features only if $\mu_\alpha/\mu\geq1$. Under this condition, we can deduce information about the quasi-classical nature of the early Universe, after reproducing a calculation similar to Misner's one \cite{QC}. In particular, we know that the point-Universe experiences a never-ending series of bounces towards the singularity and its direction changes according to \eqref{reflection}. As we mentioned above, we can infer that the particle motion has a stochastic nature. Then, we can assume that our analysis won't be influenced by the choice of the initial conditions. In particular, the choice of $\theta_i+\theta_f=60^\circ$ for the first bounce implies $\theta_i=\theta'_i$ and $\theta_f=\theta'_f$ (see Fig. \ref{fig:1}) for all the subsequent bounces and this simplify the following calculations.

\begin{Figure}[h!]
	\centering
	\includegraphics[width=0.5\linewidth]{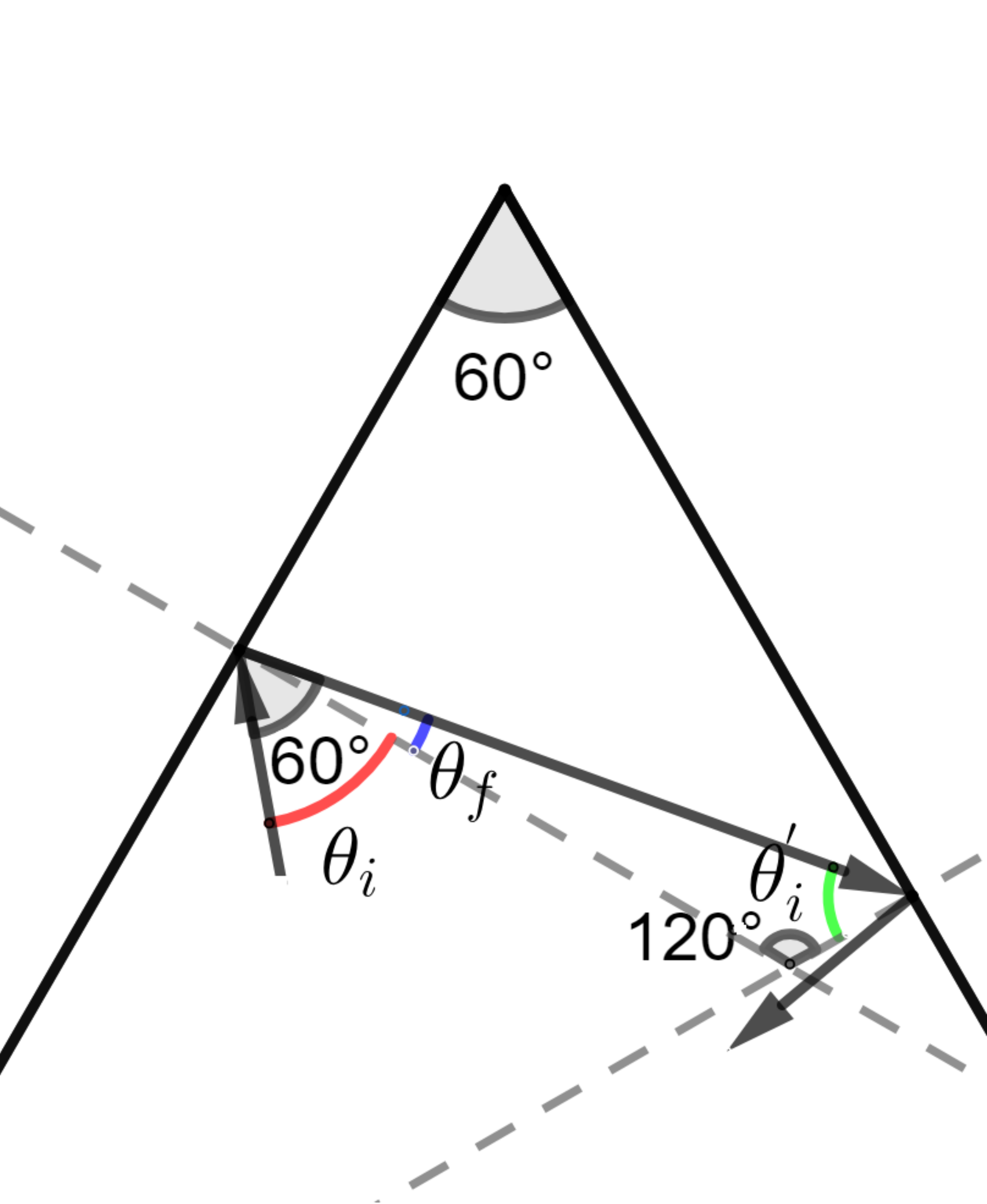}
	\captionof{figure}{\label{fig:1} Relation between the incident angles of two subsequent bounces. The figure shows that the initial condition $\theta_i+\theta_f=60^\circ$ implies $\theta_i=\theta'_i$, in fact $\theta_i'=180^\circ-120^\circ-\theta_f=180^\circ-120^\circ-(60^\circ-\theta_i)=\theta_i$.}
\end{Figure}

Therefore, using \eqref{p-}, we find the relation

\begin{equation}
\label{cost}
\frac{r_i\sin(\mu_\alpha H_{\alpha_i})\sqrt{1-\sin^2(\mu_\alpha H_{\alpha_i})}}{r_f\sin(\mu_\alpha H_{\alpha_f})\sqrt{1-\sin^2(\mu_\alpha H_{\alpha_f})}}=\frac{\sin\theta_f}{\sin\theta_i}=const\,,
\end{equation}

since $\theta_i$ and $\theta_f$ are constant. Now, if we apply the Sine theorem to the triangles $ABC$ and $ABD$ (Fig. \ref{fig:2}) we get:

\begin{eqnarray}
\label{rel1}
&&\frac{r_f|\alpha_f|}{\sin120^\circ}=\frac{|\alpha|/2}{\sin\theta_i}\,,\\
\label{rel2}
&&\frac{r_i|\alpha_i|}{\sin120^\circ}=\frac{|\alpha|/2}{\sin\theta_f}\,,
\end{eqnarray}

where we have used the initial condition $\theta_i+\theta_f=60^\circ$. 

\begin{Figure} [h!]
	\centering
	\includegraphics[width=0.85\linewidth]{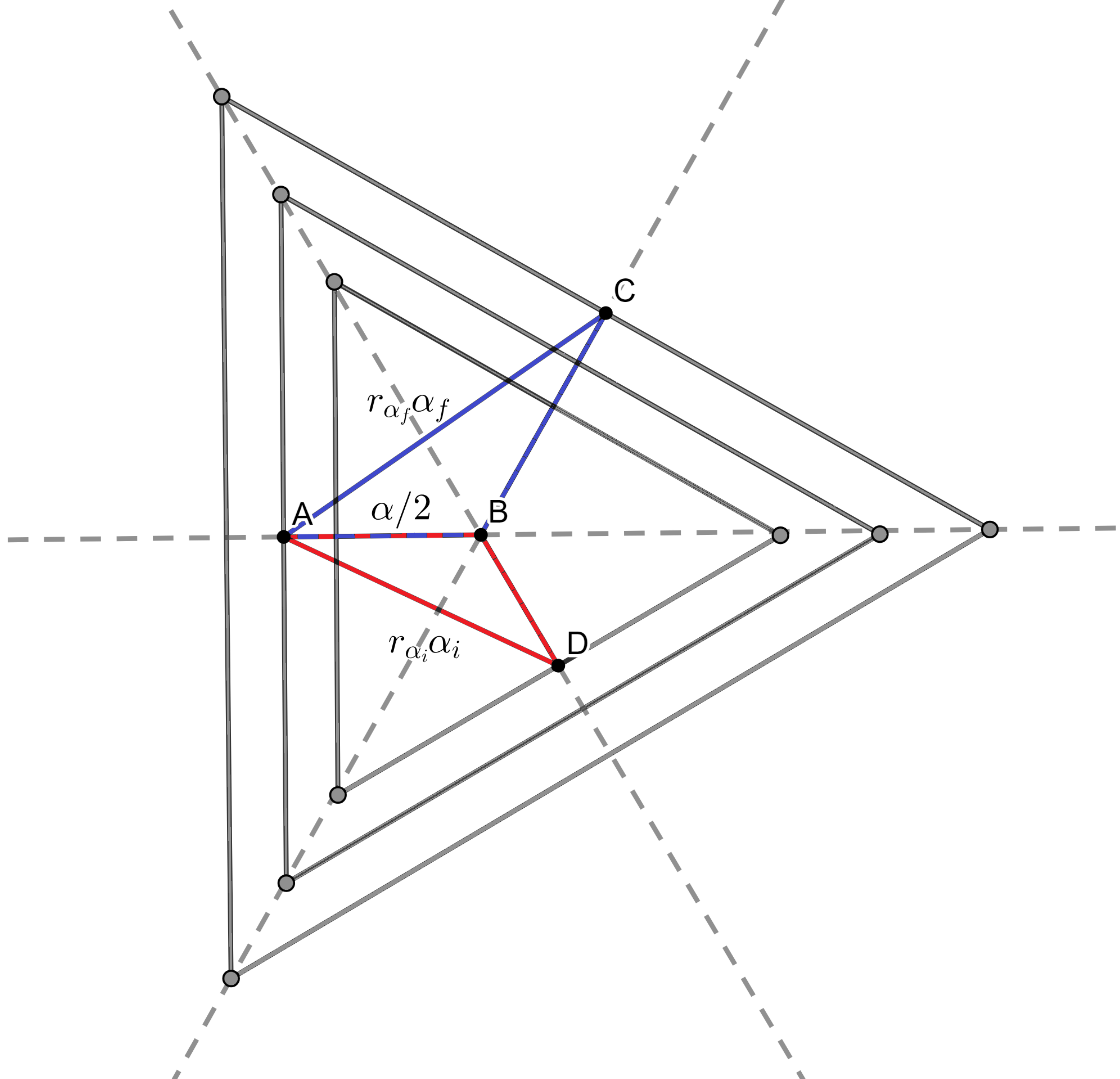}
	\captionof{figure}{\label{fig:2} Geometric relations between two subsequent collisions. In particular, here we fix $\alpha$ as the instant in which the first bounce occurs, so the position of the potential wall is $|\beta_{wall}|=|\alpha|/2$, while the distances traveled by the point-Universe before and after the bounce are respectively $d_i=r_i|\alpha_i|$ and $d_f=r_f|\alpha_f|$.}
\end{Figure}

These last two relations imply that

\begin{equation}
\label{rel3}
r_i|\alpha_i|\sin\theta_f=r_f|\alpha_f|\sin\theta_i\,,
\end{equation}

so, combining \eqref{rel3} and \eqref{cost}, we obtain a quantity that assumes the same value before and after the bounce:

\begin{multline}
\label{equation}
r_i^2|\alpha_i|\sin(\mu_\alpha H_{\alpha_i})\sqrt{1-\sin^2(\mu_\alpha H_{\alpha_i})}=\\=r_f^2|\alpha_f|\sin(\mu_\alpha H_{\alpha_f})\sqrt{1-\sin^2(\mu_\alpha H_{\alpha_f})}\,.
\end{multline}

In order to generalize this result, we consider the bounce to occur at the time $\alpha\equiv\alpha_n$ and we apply again the Sine theorem to the same triangles, finding the expressions 

\begin{eqnarray}
&&r_i|\alpha_i|=|\alpha_{n}|\frac{\sin120^\circ}{2\sin\theta_f}=C|\alpha_{n}|\,,\\
&&r_f|\alpha_f|=|\alpha_{n+1}|\frac{\sin120^\circ}{2\sin\theta_f}=C|\alpha_{n+1}|\,.
\end{eqnarray}

Substituting these relations in \eqref{equation} and taking the average value over a large number of bounces, we derive the following \emph{adiabatic invariant}:

\begin{equation}
\label{adiabatic}
\big<r|\alpha|\sin(\mu_\alpha H_{\alpha})\sqrt{1-\sin^2(\mu_\alpha H_{\alpha})}\big>=const\,.
\end{equation}

This is not a constant of motion but it represents a quantity that turns to have the same value just before each bounce.

As we are interested in the existence of quasi-classical states near the singularity, we can evaluate the quantity \eqref{adiabatic} in the quasi-classical approximations $H_\alpha\simeq\bar{p}_\alpha$, $p_+\simeq p_+^\mu$, $p_-\simeq p_-^\mu$ (see \eqref{pmu}, \eqref{chi} and \eqref{kappa}) and take the limit $\alpha\rightarrow-\infty$.
Following this procedure, we obtain the same result derived by Misner in \cite{QC}, i.e.

\begin{equation}
\big<\sqrt{m^2+n^2}\big>=const\,,
\end{equation}

which means that the occupation number is an adiabatic invariant. Therefore, we can conclude that quasi-classical states for the Mixmaster anisotropies are allowed towards the singularity even in the polymer representation, provided that the choice of the polymer parameters implies a chaotic dynamics.

\subsection{The free particle in an equilateral triangular well}

In \cite{QC} Misner implemented the Bianchi IX quantization through the observation that the eigenvalue problem was that one of a free particle in an equilateral triangular potential well, once you consider the exponential walls as infinitely steep and this is naturally reasonable in the limit $\alpha\rightarrow-\infty$. He also estimated that the eigenvalues for a triangle would be quite the same as those for a square, with the advantage that the eigenfunctions and the eigenvalues of the square problem are elementary and well-known.
In polymer-quantizing the Mixmaster model we have also approximated the triangular well with a square one of the same area in order to have the possibility to verify the consistence of our results through the comparison with those ones obtained by Misner. However, here we want to complete our study by discussing the validity of the quantum analysis in the absence of the square approximation. Indeed, the triangular problem is solved explicitly in the ordinary quantum mechanics, as shown in \cite{T1,T2}, so these results can be used to derive the eigenvalues and the eigenfunctions in the more general polymer formulation.

We start with considering an equilateral triangle with vertices located at $(-\alpha/2,-\sqrt{3}\alpha/2)$, $(\alpha,0)$, $(-\alpha/2,\sqrt{3}\alpha/2)$ in the $(\beta_+,\beta_-)$ plane. The eigenvalue problem is represented by the following Schroedinger equation:

\begin{equation}
\left[p_+^2+p_-^2+V(\alpha,\beta_\pm)\right]\phi(p_\pm)=E^2\phi(p_\pm)
\label{Hpol}
\end{equation}

where the potential term has the form\footnote{We notice that here the variable $\alpha$ is considered to be only a parameter.}

\begin{equation}
V(\alpha,\beta_\pm)=\begin{cases} 0 & -\frac{\alpha}{2}\leq\beta_+\leq\alpha\, , \,-\frac{\alpha-\beta_+}{\sqrt{3}}\leq\beta_-\leq\frac{\alpha-\beta_+}{\sqrt{3}} \\ 
\infty & elsewhere
\end{cases}
\end{equation}

The energy spectrum is given by

\begin{equation}
E^2=\frac{1}{3\alpha^2}\bigg(\frac{4\pi}{3}\bigg)^2(m^2+n^2-mn)\,,
\end{equation}

where $m,n$ are integer numbers and respect the condition $m\geq2n$\footnote{We note that $E^2\rightarrow+\infty$ for $m,n\rightarrow+\infty$ if $m\geq2n$.}. 

For the case $m=2n$ there is a single non-degenerate eigenstate for each $n$

\begin{eqnarray}
\nonumber\phi^o_{2n,n}(\beta_\pm)&&=\sqrt{\frac{8}{9\sqrt{3}\alpha^2}}\bigg[\sin\big(2A_+(\alpha-\beta_+)\big)+\\&&-2\sin\big(A_+(\alpha-\beta_+)\big)\cos\big(\sqrt{3}A_+\beta_-\big)\bigg]\quad
\end{eqnarray}

while for $m>2n$ there are two degenerate eigenstates depending on the symmetry properties

\begin{eqnarray}
\label{phitri}
\nonumber\phi^+_{m,n}(\beta_\pm)=\sqrt{\frac{16}{9\sqrt{3}\alpha^2}}\bigg[&&-\sin\big(A_+(\alpha-\beta_+)\big)\cos\big(A_-\beta_-\big)+\\\nonumber&&+\sin\big(B_+(\alpha-\beta_+)\big)\cos\big(B_-\beta_-\big)+\\&&+\sin\big(C_+(\alpha-\beta_+)\big)\cos\big(C_-\beta_-\big)\bigg]\;\;\;\;\;\;
\end{eqnarray}

\begin{eqnarray}
\nonumber\phi^-_{m,n}(\beta_\pm)=\sqrt{\frac{16}{9\sqrt{3}\alpha^2}}\bigg[&&-\sin\big(A_+(\alpha-\beta_+)\big)\sin\big(A_-\beta_-\big)+\\\nonumber&&+\sin\big(B_+(\alpha-\beta_+)\big)\sin\big(B_-\beta_-\big)+\\&&-\sin\big(C_+(\alpha-\beta_+)\big)\sin\big(C_-\beta_-\big)\bigg]\;\;\;\;\;\;\;
\end{eqnarray}

In these expressions we have done the following substitutions to simplify the notation:

\begin{equation}
\label{coeff}
\begin{cases}
A_+=\frac{2\pi n}{3\alpha} & A_-=\frac{2\pi(2m-n)}{3\sqrt{3}\alpha} \\
B_+=\frac{2\pi m}{3\alpha} & B_-=\frac{2\pi(2n-m)}{3\sqrt{3}\alpha} \\
C_+=\frac{2\pi(n-m)}{3\alpha} & C_-=\frac{2\pi(n+m)}{3\sqrt{3}\alpha} 
\end{cases}
\end{equation}

In order to write the eigenfunctions of the triangular problem in the polymer formulation we have to solve the polymer-modified Schroedinger equation, which means imposing the boundary conditions on the free-particle eigenstates. In Polymer Quantum Mechanics the Hamiltonian in \eqref{Hpol} changes its formal expression due to the substitutions \eqref{pol1} and \eqref{pol2} needed to well-define the operators $p_+$ and $p_-$, but this modification doesn't affect the expression of the free-particle eigenstates that remains the same of the ordinary quantum mechanics (see \eqref{phi}). Now, the simpler way to solve the problem is to impose the boundary conditions on a linear combination of free-particle eigenstates according to all the symmetries of the triangular problem. However, the resulting expression (for example \eqref{phitri}) is not the eigenfunction of the polymer eigenvalue problem (as it happens for the eigenfunction \eqref{phi3} of the square problem) since the polymer-Hamiltonian is not invariant anymore under the rotational symmetries of the triangular potential term just because of the substitutions \eqref{pol1} and \eqref{pol2}. In order to show it, we write the Fourier transform of $\phi^+_{m,n}(\beta_\pm)$ that has the following expression:

\begin{widetext}
	\begin{eqnarray}
	\nonumber\phi^+_{m,n}(p_\pm)=\phi^+_1(p_\pm)-\phi^+_2(p_\pm)-\phi^+_3(p_\pm)=\\\nonumber=i\sqrt{\frac{4\pi^2}{9\sqrt{3}\alpha^2}}\bigg[&&\big(e^{iA_+\alpha}\delta(p_+-A_+)-e^{-iA_+\alpha}\delta(p_++A_+)\big)\cdot\big(\delta(p_--A_-)+\delta(p_-+A_-)\big)+\\\nonumber-&&\big(e^{iB_+\alpha}\delta(p_+-B_+)-e^{-iB_+\alpha}\delta(p_++B_+)\big)\cdot\big(\delta(p_--B_-)+\delta(p_-+B_-)\big)+\\\nonumber-&&\big(e^{iC_+\alpha}\delta(p_+-C_+)-e^{-iC_+\alpha}\delta(p_++C_+)\big)\cdot\big(\delta(p_--C_-)+\delta(p_-+C_-)\big)\bigg]\quad
	\end{eqnarray}
\end{widetext}
\newpage
The action of the polymer-Hamiltonian on $\phi^+_{m,n}(p_\pm)$ is

\begin{eqnarray}
\nonumber\bigg[\frac{1}{\mu^2}\sin^2(\mu p_+)+\frac{1}{\mu^2}\sin^2(\mu p_-)\bigg]\phi^+_{m,n}(p_\pm)&&=\\=\left[E_1^2\phi^+_1(p_\pm)-E_2^2\phi^+_2(p_\pm)-E_3^2\phi^+_3(p_\pm)\right]&&\,,
\end{eqnarray}

where

\begin{equation}
\begin{cases}
E_1^2=\frac{1}{\mu^2}\sin^2(\mu A_+)+\frac{1}{\mu^2}\sin^2(\mu A_-) \\
E_2^2=\frac{1}{\mu^2}\sin^2(\mu B_+)+\frac{1}{\mu^2}\sin^2(\mu B_-) \\
E_3^2=\frac{1}{\mu^2}\sin^2(\mu C_+)+\frac{1}{\mu^2}\sin^2(\mu C_-) 
\end{cases}
\end{equation}

and this highlights the fact that $\phi^+_{m,n}(p_\pm)$ is not the eigenfunction of the triangular problem in the polymer representation. Nevertheless, in the limit $\alpha\rightarrow-\infty$ we can make a first order expansion of the sine function to show that the coefficients $E_1^2$, $E_2^2$, $E_3^2$ tends to the same expression $E^2$ that in this way can be considered to be the energy eigenvalue. In fact, using the explicit expressions in \eqref{coeff} we obtain

\begin{eqnarray}
\nonumber A_+^2+A_-^2=B_+^2+B_-^2=C_+^2+C_-^2&&=\\=\frac{1}{3\alpha^2}\bigg(\frac{4\pi}{3}\bigg)^2(m^2+n^2-mn)=E^2&&\,.
\end{eqnarray}

Since we are dealing with a limit to the cosmological singularity ($\alpha\rightarrow-\infty$) in order to approximate the potential with a real well, the function \eqref{phitri} can be considered the natural eigenstate of the triangular polymer problem. In this limit, the relation \eqref{chi} for the energy eigenvalue becomes

\begin{equation}
\tilde{p}_\alpha=\frac{1}{\mu_\alpha}\arcsin(\mu_\alpha \tilde{k})\,,
\end{equation}

with $\tilde{k}=\sqrt{\frac{1}{3\alpha^2}\big(\frac{4\pi}{3}\big)^2(m^2+n^2-mn)}$\,.

Also, the average values of the operators $\frac{1}{\mu^2}\sin^2(\mu p_+)$ and $\frac{1}{\mu^2}\sin^2(\mu p_-)$ on this eigenstate assume the same following expression:

\begin{eqnarray}
\nonumber\tilde{p}&&=\big<\frac{1}{\mu^2}\sin^2(\mu p_+)\big>=\big<\frac{1}{\mu^2}\sin^2(\mu p_-)\big>=\\&&=\frac{1}{6\alpha^2}\bigg(\frac{4\pi}{3}\bigg)^2(m^2+n^2-mn)\,.
\end{eqnarray}

Using the new semiclassical approximations $H_\alpha\simeq\tilde{p}_\alpha$, $\frac{1}{\mu^2}\sin^2(\mu p_+)\simeq\tilde{p}$, $\frac{1}{\mu^2}\sin^2(\mu p_-)\simeq\tilde{p}$ in \eqref{adiabatic} we find the new expression of the abiabatic invariant that writes as

\begin{equation}
\big<\sqrt{m^2+n^2-mn}\big>=const\,.
\end{equation}

This result still confirm the conservation of the quantum numbers $m,n$ and so the existence of quasi-classical states near the singularity even without using the square approximation.

\section{Concluding remarks} 

We analyzed the Bianchi IX Cosmology, as described in the standard Misner variables, by implementing a semiclassical and quantum treatment of the dynamics in terms of the prescriptions of the so-called polymer paradigm. In other words, we introduced a discrete structure in all the Minisuperspace variables, in order to account for cut-off physics near the cosmological singularity. 

As already clarified by the analyses in \cite{O,Chiara}, the implementation of such a cut-off paradigm is unable to remove the singularity. In fact, here we demonstrate that the singularity of Bianchi IX is still present if the configurational variables are taken in the standard Misner choice, since discretizing the logarithm of the 
natural Universe scale factor (i.e. the $\alpha$ Misner variable) means discretizing the Universe volume but without forbidding its zero value, like it is done in \cite{S} where $\alpha$ is replaced by the cubed scale factor and the singularity is removed (see also \cite{M} for a related analysis in the case of the isotropic Universe).

The main merit of this study is showing that the equivalence between the semiclassical dynamics emerging in Polymer Quantum Cosmology and Loop Quantum Cosmology depends strongly on the choice of the phase-space variables. In Loop Quantum Cosmology the choice of these variables is forced by the non-Abelian gauge structure based on the $SU(2)$ group emerging in the Ashtekar-Barbero-Immirzi connection variables. Moreover, when the semiclassical limit is taken in the Minisuperspace dynamics, the removal of the singularity can be considered to be connected with fixing a minimum area scale (see in this respect \cite{A2}). In Polymer Quantum Mechanics no privileged set of configurational variables exists, except by requiring the capability of the theory to reproduce the semiclassical features of Loop Quantum Cosmology (e.g. in \cite{M} is shown that to deal with a constant critical density for the isotropic Universe is necessary to adopt just the cubed scale factor, i.e. the space volume). Furthermore, in the Polymer approach the cut-off physics comes out from the discrete nature of the graph morphology on which the addressed configurational variables are represented, since this procedure implies the introduction of a minimal step automatically. When this step coincides with the minimal value of a geometric object (areas and volumes), like for the cubed scale factor, a Bouncing Cosmology is expected to emerge and the equivalence with Loop Quantum Cosmology is reliably guaranteed. The choice of the variable $\alpha$ is on a different footing, considering that the polymer discretization does not prevent the variable $\alpha$ from going to minus infinity, with the consequence that the cosmological singularity survives. In other words, in this case we are dealing with a discrete Universe volume which can also reach the zero value in the semiclassical dynamics. This situation suggests an intriguing implication regarding the possibility to have also in Loop Quantum Cosmology the zero eigenvalue in the volume spectrum and points out that a more subtle implementation of the kinematical spectrum of the space volume on a dynamical level is needed. In fact, this feature of the general theory is hidden in the Minisuperspace applications, especially when the simmetry simplifications make a bit obscure the role played by the zero eigenvalue for the volume when it is weighted on a semiclassical configuration. For related criticisms to Loop Quantum Cosmology in view of reproducing the full space graph morphology see \cite{Cianfrani1,Cianfrani2,Alesci,C}.

Concerning the study of the chaotic properties of the model, the most valuable result here achieved consists of demonstrating that the semiclassical chaotic features of the Bianchi IX Cosmology survive only if the ratio of the polymer scale parameter for the isotropic variable to that one for the anisotropies is greater than or equal to the unity. This fact proves that the chaos survives only when the Universe volume discretization is more relevant with respect to the corresponding discretization of the anisotropies, with the consequence that the main features of the Misner original description \cite{QC,MU} are preserved. On the contrary, a significant discretization of the anisotropies (i.e. the physical degrees of freedom of the cosmological gravitational field) with respect to the volume component leads to a suppression of the chaos, since new features of the Kasner solution appear.

Finally, for the parameter space associated to the chaotic features of the model, we reproduced in the polymer paradigm all the considerations made by Misner on 
a quantum level, recovering his result about the possibility to have high occupation numbers for the anisotropy
degrees of freedom, i.e. quasi-classical states, near enough to the singularity. In particular, the possibility to have a quasi-classical singular Universe appears in the considered scenario and it is unaffected by the discretization of the variables.

In conclusion, the fact that all the semiclassical and quantum features of the usual Bianchi IX model are recovered when the chaos is preserved suggests that the polymer lattice scale seems to be essentially a negligible effect with respect to the physical content of the Misner representation, except when it introduces a predominant discretization for the anisotropy degrees of freedom over that one for the Universe volume.

\end{multicols}
\end{document}